\documentclass[12pt]{iopart}
\usepackage{iopams} 
\usepackage{graphicx}
\newcommand{\bra}{\langle}
\newcommand{\ket}{\rangle}
\newcommand{\nn}{\nonumber}
\renewcommand{\i}{{\rm{i}}}
\renewcommand{\e}{{\rm{e}}}
\begin{document}
\newcommand{\bv}[1]{\mbox{\boldmath$#1$}}

\title[Creation
of Vortices in BEC]{Continuous Creation
of a Vortex in a Bose-Einstein Condensate with Hyperfine Spin $F=2$}

\author{Mikko M\"ott\"onen$^1$, Naoki Matsumoto$^2$, Mikio Nakahara$^{1,2}$
\footnote[3]{To
whom correspondence should be addressed (nakahara@math.kindai.ac.jp)}
and Tetsuo Ohmi$^3$
}
\address{
Materials Physics Laboratory, Helsinki University of Technology,
P.O. Box 2200 FIN-02015 HUT, Finalnd$^1$}
\address{
Department of Physics, Kinki University, Higashi-Osaka 577-8502, Japan$^2$}
\address{Department of Physics, Kyoto University, Kyoto 606-8502, Japan$^3$}
\begin{abstract}
It is shown that a vortex can be
continuously created in a Bose-Einstein condensate
with hyperfine spin $F=2$ in a Ioffe-Pritchard trap
by reversing the axial magnetic field
adiabatically. It may be speculated that the condensate cannot
be confined in the trap since the weak-field seeking state
makes transitions to the neutral and the strong-field seeking states
due to the degeneracy of these states along
the vortex axis when the axial field vanishes. We have solved the 
Gross-Pitaevskii equation numerically with given external
magnetic fields to show that this is not the case.
It is shown that a considerable fraction of the condensate remains in the
trap even when the axial field is reversed rather slowly.
This scenario is also analysed in the presence of an optical plug
along the vortex axis. Then the condensate remains within
the $F_z=2$ manifold, with respect to the local magnetic field,
throughout the formation of a vortex
and hence the loss of atoms does not take place. 
\end{abstract}



\maketitle

\section{Introduction}

It has been observed that the Bose-Einstein condensate (BEC)
of alkali atom gas becomes superfluid \cite{rf:1, rf:2}. Superfluidity
of this system is different from the previously known
superfluid $^4$He in many aspects. For example, the BEC is
a weak-coupling gas for which the Gross-Pitaevskii equation
is applicable while superfluid $^4$He is a strong-coupling system.
One of the most remarkable differences is that
the BEC has spin degree of freedom
originating from the hyperfine spin of the atom, and that this degree of
freedom couples to external magnetic fields.
Accordingly the order parameter of the condensate is also
controlled at will by external magnetic fields. 
Superfluid $^3$He also has similar internal degrees of freedom,
which, however, are rather difficult to control by external fields \cite{rf:3}.

Taking advantage of this observation, we proposed a simple method
to create a vortex in a BEC with the hyperfine spin $F=1$ \cite{tomoya, ogawa};
a vortex-free BEC is intertwined topologically by manipulating the magnetic
fields in the Ioffe-Pritchard trap to form a vortex with the winding
number 2. This is achieved by reversing the axial magnetic field adiabatically
while the planar quadrupole field is kept fixed. 

In the present paper, a similar scenario is analysed for
a BEC with $F=2$, taking $^{87}$Rb as an example.
The difference between the present case and that for $F=1$
will be emphasised in our analysis. 
In the next section, we briefly review the order parameter of
BEC with hyperfine spin $F=2$ and the Gross-Pitaevskii (GP) equation which
describes the time-evolution of the order parameter.
In section 3,
the ground state order parameter of the BEC in the weak-field seeking state
confined in a harmonic potential is obtained. 
Then the time-evolution of the condensate, as the axial field
is adiabatically reversed, is studied by solving the GP equation
numerically. Cases with different reverse time are analysed
to find the best possible reversing time for which
the the fraction of the remaining condensate in the vortex state is maximised.
It is shown that the condensate in the end of this scenario has
the winding number 4. In section 4,
the GP equation is solved in the presence of an optical plug along
the vortex axis. The BEC remains in the
$F_z=2$ weak-field seeking state, with respect to the local
magnetic field, throughout the development,
and hence no atoms will be lost during the formation of a vortex.
Section 5 is devoted to conclusions and discussions.

\section{Order Parameter of $F=2$ BEC}

\subsection{General $F=2$ condensate and Gross-Pitaevskii equation}

Suppose a uniform magnetic field $\bv{B}$ parallel to the $z$-axis is applied 
to a BEC of alkali atoms with the hyperfine spin $F=2$.
Then the hyperfine spin state of the atom is quantised along
this axis; the eigenvalue $m$ of $F_z$ takes a value $-2 \leq m \leq +2$,
where $F_z | m \ket = m |m \ket$.
Let us introduce the following conventions
$$
|2 \ket = \left(
   \begin{array}{c}
      1\\ 0\\ 0\\ 0\\ 0
   \end{array} \right),
   \ |1 \ket = \left(
   \begin{array}{c}
      0\\ 1\\ 0\\ 0\\ 0
   \end{array} \right),
   \ |0 \ket = \left(
   \begin{array}{c}
      0\\ 0\\ 1\\ 0\\ 0
   \end{array} \right),
   \ |-1 \ket = \left(
   \begin{array}{c}
      0\\ 0\\ 0\\ 1\\ 0
   \end{array} \right),
   \ |-2 \ket =
   \left( \begin{array}{c}
      0\\ 0\\ 0\\ 0\\ 1
   \end{array} \right).
$$
The order parameter $|\Psi \ket$ is expanded in terms of $|m \ket$ as
\begin{equation}
|\Psi \ket = \sum_{m=-2}^2 \Psi_m |m \ket =
(\Psi_2,\Psi_1,\Psi_0,\Psi_{-1},\Psi_{-2})^{\rm T},
\end{equation}
where T denotes the transpose.

The representation of the angular momentum operators $F_k\ (k=x, y, z)$ for $F=2$
is easily obtained from the well-known formulae
\begin{eqnarray*}
\bra 2, m|F_+|2, m' \ket &=& \sqrt{(2-m)(3+m)} \delta_{m, m'+1},\\
\bra 2, m|F_-|2, m' \ket &=& \sqrt{(2+m)(3-m)} \delta_{m, m'-1},\\
\bra 2, m|F_z|2, m' \ket &=& m \delta_{m, m'},
\end{eqnarray*}
where $F_{\pm} = F_x \pm i F_y$.

The dynamics of the condensate in the limit of zero temperature is given,
within the mean field approximation, by the time-dependent
Gross-Pitaevskii (GP) equation with spin degrees of freedom.
This equation, obtained by Ciobanu, Yip and Ho \cite{Ciobanu}
(see also \cite{ueda}),
is written in components $\Psi_m$ as
\begin{eqnarray}\label{eq:gptime}
   i\hbar\frac{\partial}{\partial t}\Psi_{m} &=& \left[-\frac{\hbar^2}{2M}
   \nabla^2 +V(r) \right] \Psi_m \nn\\
& & +g_1|\Psi_n|^2 \Psi_m 
   + g_2\left[\Psi_{n}^\dagger (F_{k})_{np} \Psi_{p}\right]
   (F_{k})_{mq}\Psi_q \nn\\
   & &+ 5g_3\Psi_n^\dagger \bra 2m2n|00\ket\bra 00|2p2q\ket\Psi_p\Psi_q 
   +\frac{1}{2}\hbar\omega_{Lk}(F_k)_{mn}\Psi_n,
\end{eqnarray}
where summations over $k=x,y,z$ and $-2 \leq n,p,q \leq 2$ are understood.
Here, $M$ is the mass of the atom and $V(r)$ is
the possible external potential. 
The Larmor frequency is defined as $\hbar\omega_{Lk}=\gamma_\mu B_k$, where
$\gamma_\mu\quad \lsimeq \mu_B$ is the gyromagnetic ratio of the atom and
$\mu_B$ the Bohr magneton. The interaction parameters are
expressed in terms of the $s$-wave
scattering length $a_F$, $F$ being the total
hyperfine spin of the two-body scattering state, and are given by
\cite{Ciobanu}
\begin{eqnarray}
g_1 = \frac{4\pi\hbar^2}{M}\frac{4a_2+3a_4}{7} \nn \\
g_2 = \frac{4\pi\hbar^2}{M}\frac{a_2-a_4}{7} \nn \\
g_3 = \frac{4\pi\hbar^2}{M}\left(\frac{a_0-a_4}{5}-\frac{2a_2-2a_4}{7}\right),
\end{eqnarray}
where $a_0=4.73$nm, $a_2=5.00$nm and $a_4=5.61$nm for $^{87}$Rb atoms.
This should be compared with $F=1$ BEC where there are only two types
of scattering state and hence two interaction terms in the GP equation.

\subsection{Weak-field seeking state}

Suppose a strong magnetic field $\bv{B}$ is applied along the $z$-axis.
Then the components with $F_z = 1$ and $2$ are in the weak-field seeking state
(WFSS)
while those with $F_z = -1$ and $-2$ are in the strong-field seeking
state (SFSS). The presence of two WFSSs leads to an interesting
two-component vortex that is not observed in $F=1$ BEC as we see in
the next section.
The energy of the state with $F_z=0$ is independent of the magnetic field
and will be called the neutral state (NS) hereafter.
Suppose a uniform condensate is in the state with $F_z=2$.
The order parameter of the condensate takes the form
\begin{equation}
|\Psi_0 \ket = f_0 \left( 1, 0, 0, 0, 0
\right)^{\rm T}
\label{eq:b0}
\end{equation}
where $|f_0|^2$ is the number density of the condensate. Now let us consider
a state which is quantised along an arbitrary local magnetic field
\begin{equation}
\bv{B}(\mathbf{r})= B \left( \begin{array}{c}
\sin \beta \cos \alpha\\
\sin \beta \sin \alpha\\
\cos \beta
\end{array} \right).
\label{eq:mag}
\end{equation}
Let $F_B \equiv \bv{B} \cdot \bv{F}/B$
be the projection of the hyperfine
spin vector along the local magnetic field.
The WFSS $|\Psi \ket$ which satisfies $F_B |\Psi \ket
= +2 |\Psi \ket$ is obtained by rotating $|\Psi_0 \ket$ by
Euler angles $\alpha, \beta$ and $\gamma$ and is given by
\begin{eqnarray}
|\Psi (\mathbf{r}) \ket&=&
\exp(-i \alpha F_z) \exp(-\beta F_y) \exp(-i\gamma F_z) |\Psi_0 \ket
\nonumber\\
&= &f_0 e^{-2 i\gamma} \left( \begin{array}{c}
e^{-2i \alpha} \cos^4 \frac{\beta}{2}\\
2 e^{-i \alpha} \cos^3 \frac{\beta}{2}\sin \frac{\beta}{2}\\
\sqrt{6} \cos^2 \frac{\beta}{2}\sin^2 \frac{\beta}{2}\\
2 e^{i \alpha} \cos \frac{\beta}{2}\sin^3 \frac{\beta}{2}\\
e^{2i \alpha} \sin^4 \frac{\beta}{2}
\end{array}
\right) \equiv f_0 |v \ket.
\label{eq:wfss}
\end{eqnarray}

The GP equation restricted within the WFSS is obtained by substituting
Eq.~(\ref{eq:wfss}) into Eq.~(\ref{eq:gptime}), see the next section.

\section{Vortex formation without optical plug}

The formation of a vortex in the $F=2$ condensate is analysed in this and the
next sections. In the present section, we study the scenario without an
optical plug along the vortex axis. Although some fraction of the condensate
is lost from the trap in this scenario, the experimental setup will be much easier
without introducing an optical plug. In fact, it will be shown below that
a considerable amount of the condensate remains in the trap by
properly choosing the time-dependence of the magnetic field.

\subsection{Magnetic fields}

Suppose a condensate is confined in a Ioffe-Pritchard trap.
It is assumed that the trap is translationally invariant along
the $z$ direction and rotationally invariant around the $z$-axis.
The quadrupole magnetic field of the trap
takes the form
\begin{equation}
\bv{B}_{\perp} (\bv{r}) = B_{\perp}(r) \left( \begin{array}{c}
\cos (-\phi)\\
\sin (-\phi)\\
0
\end{array}
\right),
\end{equation}
where $\phi$ is the polar angle.
The magnitude $B_{\perp}(r)$ is proportional to the radial distance
$r$ near the origin; $B_{\perp}(r) \sim B_{\perp}' r$. 
The uniform time-dependent field
\begin{equation}
\bv{B}_z(t) = \left( \begin{array}{c}
0\\
0\\
B_z(t)
\end{array}
\right)
\end{equation}
is also applied along the $z$-axis to prevent
Majorana flips from taking place at $r \sim 0$
where $\bv{B}_{\perp}$ vanishes.
Now the total magnetic field is given by
\begin{equation}
\bv{B}(\bv{r}, t) = \bv{B}_{\perp}(\bv{r}) + \bv{B}_z(t)
= \left( \begin{array}{c}
B_{\perp}(r) \cos (-\phi)\\
B_{\perp}(r) \sin (-\phi)\\
B_z(t)
\end{array}
\right).
\label{eq:magx}
\end{equation}
Comparing this equation with Eq. (\ref{eq:mag}), it is found that
\begin{equation}
\alpha = -\phi \qquad \beta = \tan^{-1}\left[\frac{B_{\perp}(r)}{B_z(t)}
\right].
\label{eq:phi}
\end{equation}

It has been shown in the previous works for $F=1$ BEC that a 
vortex-free condensate in the beginning will end up with a
condensate with a vortex of the winding number 2 if $\bv{B}_z$
reverses its direction while $\bv{B}_{\perp}$ is kept unchanged \cite{tomoya,
ogawa}.
We expect that the same magnetic field manipulation to lead
to the vortex formation in a BEC with $F=2$.
The uniform axial field $\bv{B}_z(t)$ must reverse its direction
as
\begin{equation}
B_z(t) = \left\{ 
\begin{array}{cl}
B_z(0) \left(1-\frac{2t}{T}\right)& 0 \leq t \leq T\\
-B_z(0)& T < t.
\end{array} \right.
\end{equation}
to create a vortex along the $z$-axis. This gives a ``twist'' to
the condensate leading to the formation of a vortex with winding
number 4, see below.

Before we start the detailed analysis, it will be useful
to outline the idea underlying our scenario.
Suppose one has WFSS with $F_B = 2$
in the trap at $t=0$. The magnetic field at $r \sim 0$
points $+z$ direction (i.e., $\beta \sim 0$)
and hence the WFSS takes the form $\Psi_0$
of Eq.~(\ref{eq:b0}). Then the angle $\gamma$ must satisfy
$\gamma = -\alpha$ for the BEC to be vortex-free, see Eq.~(\ref{eq:wfss}).
The field $B_z(t)$ vanishes at $t =T/2$, for which $\beta = \pi/2$,
and the hyperfine spin is parallel to the quadrupole field $\bv{B}_{\perp}$.
Accordingly one must choose $\alpha = -\phi$ for this condition to
be satisfied, see Eqs.~(\ref{eq:magx}) and (\ref{eq:phi}).
This also implies $\gamma = + \phi$.
When the field $B_z$ is completely reversed
at $t=T$, the magnetic field at $r \sim 0$ points down and hence
$\beta \sim \pi$ there. Substituting $\alpha= -\gamma =\phi$ and
$\beta = \pi$ into Eq.~(\ref{eq:wfss}), one finds the
order parameter at $t=T$;
\begin{equation}
|\Psi \ket = f_0 \e^{-2\i \phi} \left(
0, 0,0,0,\e^{-2\i \phi}
\right)^{\rm T},
\end{equation}
which shows that a vortex with the winding number 4 has been created.

\subsection{Initial state}

Suppose a vortex-free BEC is confined in a Ioff-Pritchard trap,
whose magnetic field takes the form (\ref{eq:magx}) and that
the condensate is in the eigenstate $F_B =2$ with respect
to the local magnetic field $\bv{B}$ with $B_z=B_z(t=0)$.
The condensate wave function is then obtained by solving the
stationary state Gross-Pitaevskii equation. Substitution of
Eq.~(\ref{eq:wfss}) with $\alpha = -\gamma = -\phi$ into
Eq.~(\ref{eq:gptime}) yields
\begin{eqnarray*}
-\frac{\hbar^2}{2M} \nabla^2(f_0 v_m) +(g_1+ 4 g_2) f_0^3 v_m
+ \hbar \omega_L f_0 v_m = \mu f_0 v_m,
\end{eqnarray*}
where we have put $\Psi_m \equiv f_0 v_m$. 
Note that the $g_3$ term vanishes identically for the present state.
The condensate wave amplitude
$f_0(r)$ is taken to be a real function without loss of generality.
The eigenvalue $\mu$ is identified with the chemical potential.
If one multiplies the above equation by $\{v_m\}^{\dagger}$ from the left and uses the
identity $\sum_m|v_m|^2 =1$ and other identities derived from this,
one obtains the reduced GP equation for $f_0(r)$;
\begin{equation}
-\frac{\hbar^2}{2M} \left[f_0'' +\frac{f_0'}{r} + (v_m^* \nabla^2 v_m) f_0
\right]
 + (g_1 + 4 g_2) f_0^3 + \hbar \omega_L f_0 = \mu f_0,
\end{equation}
where
\begin{equation}
v_m^* \nabla^2 v_m = \left[\frac{2}{r^2}(3 \cos^2 \beta - 5) \sin^2
\frac{\beta}{2} - \beta'^2 \right]
\end{equation}
comes from the rotation of the five-dimensional local orthonormal frame
that defines the order parameter.
The reduced GP equation looks similar to the ordinary scalar GP
equation except that there is an extra term $\beta'^2$
in $v_m^* \nabla^2 v_m $.

It is convenient to introduce the energy scale $\hbar \omega$
and the length scale $a_{\rm HO}$ define by
\begin{equation}
\omega = \sqrt{\frac{\gamma_{\mu}}{M B_z(0)} }B_{\perp}'
\qquad
a_{\rm HO} = \sqrt{\frac{\hbar}{M \omega}}.
\end{equation}
For a typical values $B_z(0) = 1{\rm G}, B_{\perp}' = 200{\rm G/cm}$
for $^{87}$Rb, one obtains $\hbar \omega \simeq 1.69\cdot10^{-24} \rm{erg}$
and $a_{\rm HO} \simeq
0.68 \mu\rm{m}$. After scaling all the physical quantities by these units, one
obtains the dimensionless form of the reduced GP equation;
\begin{eqnarray}
\lefteqn{
-\frac{1}{2} \left[\tilde{f}_0'' +\frac{\tilde{f}_0'}{\tilde{r}} +
\left[\frac{2}{\tilde{r}^2}(3 \cos^2 \beta - 5) \sin^2
\frac{\beta}{2} - \beta'^2 \right]
 \tilde{f}_0\right]}
\nn\\
& & + (\tilde{g}_1 + 4 \tilde{g}_2)
\tilde{f}_0^3 + \tilde{\omega}_L \tilde{f}_0 = 
\tilde{\mu} \tilde{f}_0,
\end{eqnarray}
where the dimensionless quantities are denoted with tilde. For example,
$\tilde{r} = r/a_{\rm HO}, \tilde{f}_0 = f_0 a_{\rm HO}^{3/2}$ and
$\tilde{g}_k = g_k/(\hbar\omega a_{\rm HO}^3)$. The tilde will be dropped hereafter 
unless otherwise stated explicitly. 
\begin{figure}
\begin{center}
\includegraphics[width=7cm]{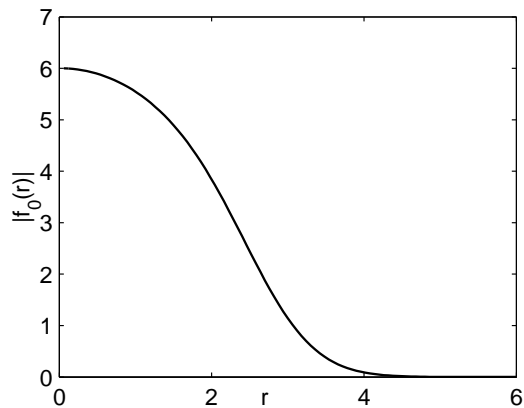}
\end{center}
\caption{\label{fig1}The initial
condensate wave function $f_0(r)$ in dimensionless
form. The radial coordinate $r$ is also dimensionless.}
\end{figure}

Figure 1 shows the ground state condensate wave function
obtained by solving Eq.~(16) numerically. 
We have chosen $f_0(r=0) = 6$
which yields the central density $n_0 \sim 1.17 \cdot 10^{14}{\rm cm}^{-3}$.
This is roughly of the same order as that realised experimentally.
The difference between the 
chemical potential and the Larmor energy at the origin is 
$\delta \mu = \mu - \omega_L = 3.95$, which amounts to
$\delta\mu = 6.68 \cdot 10^{-24} {\rm erg}$
in dimensionful form. 

\subsection{Time development}

Now the time-dependent GP equation (\ref{eq:gptime}) is solved numerically with
the initial condition $\Psi_m = f_0 (r) v_m$ with $f_0$
been obtained in the previous subsection. 
We have introduced a tanh-shaped cutoff to mimic the loss of atoms
from the trap;
particles reach at $L \gg a_{\rm HO}$ vanish from the system.
We have made several choices of the reversing time $T$ and
maximised 
the fraction of the condensate left in the trap in the final
equilibrium state. The details of the algorithm are given in 
\cite{ogawa} and will not be repeated here. 

\begin{figure}
\begin{center}
\includegraphics[width=11cm]{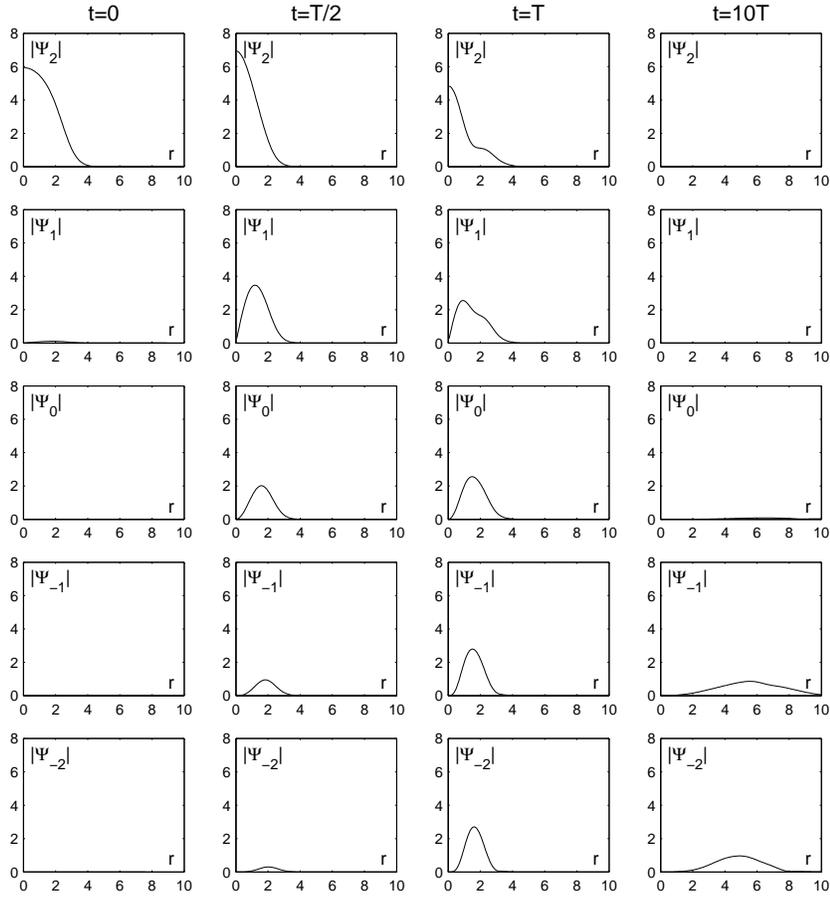}
\end{center}
\caption{Time dependence of the order parameter
$|\Psi_m|$ for the reversing time $T/\tau = 1000$.}
\end{figure}
Figure 2 shows the wave functions $|\Psi_m|$ for the choice
$T/\tau = 1000$, where $\tau = 2\pi/\omega_L$ is the time scale
set by the Larmor frequency at $t=0$ and $r=0$. 
One obtains $\tau \sim 7.14 \cdot 10^{-7} {\rm sec}$ for the
parameters given in the previous subsection.
The parameter $\tau$
is expected to be the measure of the adiabaticity. 
There are two weak-field seeking states possible for $F=2$, those
with $F_B = 2$ and $F_B=1$. It turns out that the final
vortex state is a mixture of these two states.
When the axial field $B_z(t)$ vanishes at $t=T/2$, the 
gaps among WFSSs, SFSSs and NS disappear at $r=0$ and the
level crossing takes place there. Then the adiabatic
assumption breaks down and some fraction of the condensate
transforms into SFSSs and NS as well as $F_B=1$ WFSS. 
Those components in SFSSs and NS eventually leave the trap
and the final condensate is made of $F_B=2$ and $F_B=1$ components.
It is a remarkable feature of the $F=2$ BEC, compared to
its $F=1$ counterpart, that the vortex state thus created is
mixture of these two WFSSs.
The composite nature of the final vortex state is best revealed
by projecting $|\Psi(\bv{r}) \ket$ to $F_B = 1$ and $2$ states.
Let $|v \ket$ be the vector defined in Eq.~(\ref{eq:wfss})
and $|u \ket = \exp(-i \alpha F_z) \exp(-\beta F_y) \exp(-i\gamma F_z)|1 \ket$.
Then $\Pi_2(\bv{r}) \equiv \bra v(\bv{r})|\Psi(\bv{r}) \ket$ and
$\Pi_1(\bv{r}) \equiv \bra u(\bv{r})|\Psi(\bv{r}) \ket$
depict
the projected amplitudes
of $|\Psi(\bv{r}) \ket$ to the local $F_B = 2$ and $F_B=1$ state,
respectively. These amplitudes are shown in Fig.~3 for $|\Psi(\bv{r})
\ket$ at $t= 10 T$. It is interesting to note that the $F_B=2$ component has 
a winding number 4 while $F_B=1$ has 3.
\begin{figure}
\begin{center}
\includegraphics[width=10cm]{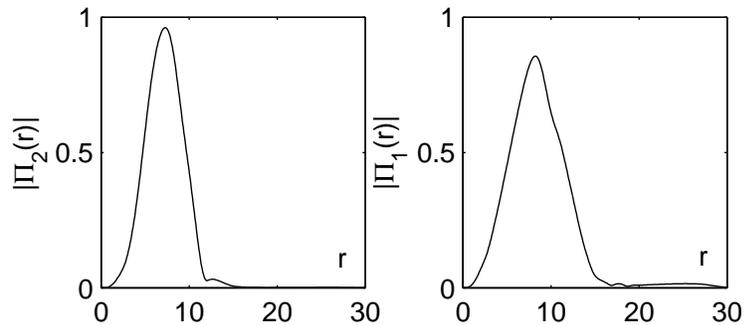}
\end{center}
\caption{The projected amplitudes $|\Pi_2(r)|$ and $|\Pi_1(r)|$
obtained from the order parameter $|\Psi(r)\ket$ at $t=10T$.}
\end{figure}

The fraction of the condensate
left in the trap at time $t$ has been plotted in Fig.~4
for $T/\tau = 1000$. It should be noted that $\sim 2/5$
of the condensate is left in the trap when the system reaches
an equilibrium at $t \gg T$.
\begin{figure}
\begin{center}
\includegraphics[width=7cm]{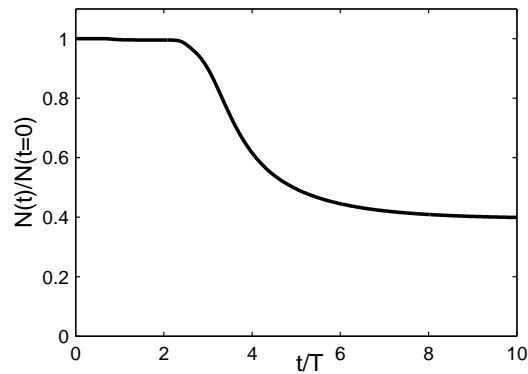}
\end{center}
\caption{The fraction of the condensate left in the
trap, as a function of the dimensionless time $t/\tau$,
for the reversing time $T/\tau = 1000$.}
\end{figure}

Figure 5 shows the fraction of the condensate left in the trap
in the equilibrium state at $t \gg T$ for various $T$.
\begin{figure}
\begin{center}
\includegraphics[width=7cm]{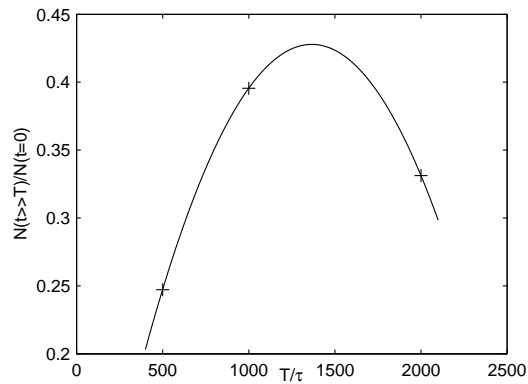}
\end{center}
\caption{\label{fig1}The fraction of the condensate left in the
trap, as a function of $T/\tau$, when the BEC reaches equilibrium
at $t \gg T$. The curve is shown for guide.}
\end{figure}
It can be seen from this figure that a considerable amount of the condensate
is left in the trap for a wide variety of the reversing time $T$.

In the next section, we analyse the creation of a vortex in
the presence of an optical plug along the centre of the system.
It will be shown that the vortex thus created is purely made
of $F_B = 2$ WFSS.

\section{Vortex formation with optical plug}

The loss of the condensate in the previous section takes place
since the energy gaps among WFSSs, NS and SFSSs disappear
at $r=0$ when $B_z$ vanishes at $t=T/2$. One may introduce
an optical plug along the vortex axis to prevent the condensate
from entering this ``dangerous'' region. An optical plug
may be simulated by a repulsive potential
\begin{equation} 
V(r) = V_0 \exp\left(-\frac{r^2}{r_0^2}\right)
\end{equation}
where $V_0$ is determined by the power of the blue-detuned laser
while $r_0$ by its waste size. We take $V_0 =
9.27 \cdot 10^{-21}\ {\rm erg}$ and $r_0 = 5 {\rm \mu m}$
in our computation below.

Now the time-independent GP equation is given by
\begin{eqnarray}
\lefteqn{
-\frac{1}{2} \left[\tilde{f}_0'' +\frac{\tilde{f}_0'}{\tilde{r}} +
\left[\frac{2}{\tilde{r}^2}(3 \cos^2 \beta - 5) \sin^2
\frac{\beta}{2} - \beta'^2 +\tilde{V}(r)\right]
 \tilde{f}_0\right]
 }
\nn\\
& & \qquad+ (\tilde{g}_1 + 4 \tilde{g}_2)
\tilde{f}_0^3 + \tilde{\omega}_L \tilde{f}_0 = 
\tilde{\mu} \tilde{f}_0
\end{eqnarray}
in dimensionless form,
where $\tilde{V}(r) = V(r)/\hbar \omega$. 
The angle $\beta$ is given by $\beta(r) = \tan^{-1}[B_{\perp}(r)/B_z(0)]$.
We will drop tilde from
dimensionless quantities hereafter unless otherwise stated.
The ground state
condensate wave function is obtained by solving this equation
numerically. We find the relative chemical potential $\delta \mu =
\mu - \omega_L = 173$, which amounts to $\delta \mu = 2.92\cdot 10^{-22} {\rm
erg}$ in dimensionful form, and the condensate wave function $f_0$
shown in Fig. 5. 

\begin{figure}
\begin{center}
\includegraphics[width=8cm]{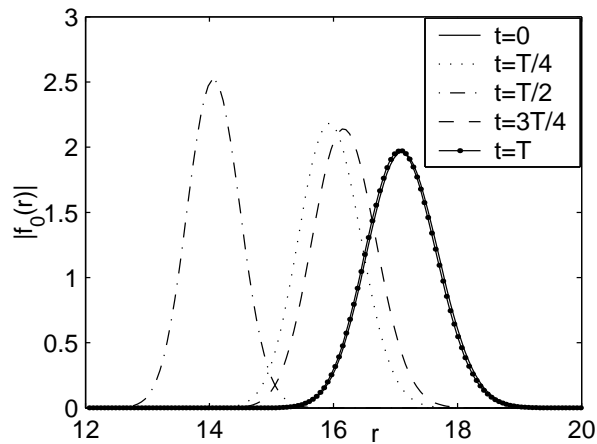}
\end{center}
\caption{Time dependence of the condensate amplitude
$f_0$ in the presence of the
optical plug. The reversing time is $T/\tau = 10000$.
The condensate amplitudes at $t=0$ and $t=T$ are almost
degenerate.
}
\end{figure}
The time-dependent GP equation
\begin{eqnarray}
\lefteqn{i\frac{\partial f_0}{\partial t}=
-\frac{1}{2} \left[{f}_0'' +\frac{{f}_0'}{{r}} +
\left[\frac{2}{{r}^2}(3 \cos^2 \beta - 5) \sin^2
\frac{\beta}{2} - \beta'^2 +{V}(r)\right]
{f}_0\right]
 }
\nn\\
& & \qquad+ ({g}_1 + 4 {g}_2) {f}_0^3 + {\omega}_L {f}_0 
\end{eqnarray}
is solved with the initial wave
function obtained above. 
Here $\beta = \beta(r, t) \equiv \tan^{-1}[B_{\perp}(r)/B_z(r)]$.
It should be stressed again that the
condensate remains within the $F_B=+2$ WFSS throughout the
temporal evolution. Figure 7 shows the time-dependence of
the components $\Psi_m$
for the choice $T/\tau = 10000$, namely $T = 7.14 {\rm ms}$
in dimensionful form.
In contrast with the case without optical plug, the time-dependence
of the order parameter is independent of the choice of $T$
so far as $T/\tau$ is large enough so that the adiabaticity is
observed. 

The vortex thus obtained has a region near the origin ($r \sim 0$)
where the condensate cannot approach due to the presence of the
optical plug. The vortex current flows around a multiply connected
region. This situation is analogous to the superconducting
current flowing around a ring.
It is natural to expect that a vortex without the
optical plug may be obtained if one withdraws the optical plug
after the persistent current is established at $t=T$.
(Note that the optical plug has been introduced to prevent
Majorana flips at $r \sim 0$ at $t \sim T/2$. Accordingly the optical
plug is not required anymore for $t \geq T$.)
Let us suppose that the optical plug is slowly turned off
after $t=T$ with the time constant $t_0$;
\begin{equation}
V(r, t) = \left\{ \begin{array}{cl}
V_0 \exp (-r^2/r_0^2)& 0< t <T\\
V_0 \exp (-r^2/r_0^2) \exp [-(t-T)/t_0]& T < t.
\end{array} \right.
\end{equation}
It is found that the condensate oscillates back and forth
for small $t_0$. For sufficiently large $t_0$, however, the
condensate smoothly rearranges itself to a vortex state without
the optical plug.
Figure 7 shows our numerical result for $T/\tau = t_0/\tau = 10000$,
for which one still observes such oscillations. 
\begin{figure}
\begin{center}
\includegraphics[width=8cm]{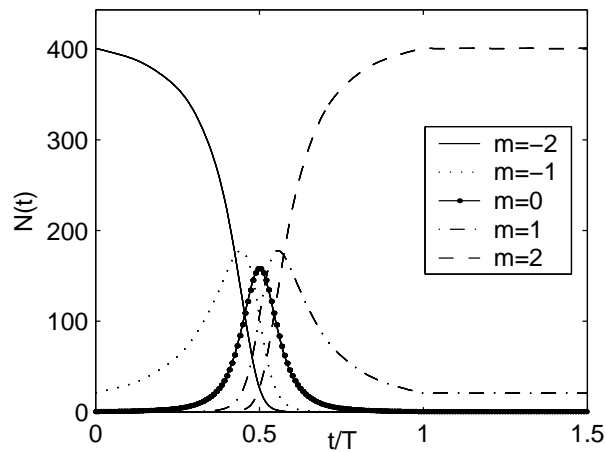}
\end{center}
\caption{The time dependence of the particle numbers in unit length 
$N_m(t) = 2\pi\int |\Psi_m(r, t)|^2 r dr$ for $T/\tau = 10000$.}
\end{figure}

\begin{figure}
\begin{center}
\includegraphics[width=8cm]{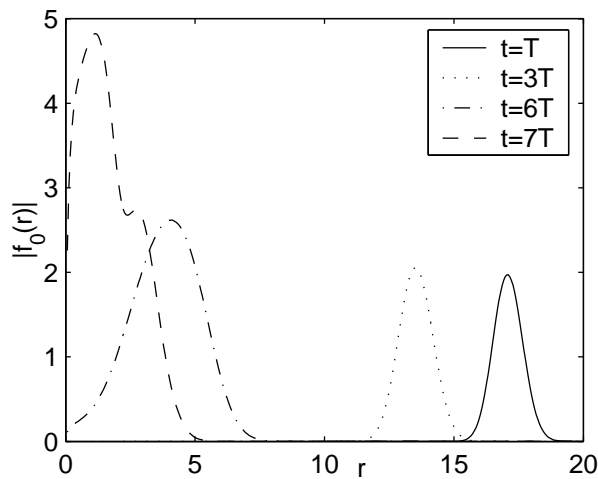}
\end{center}
\caption{Time dependence of the condensate amplitude
$f_0$ for the reversing time $T/\tau = 10000$. The potential decays 
exponentially with the time constant $t_0 = T$ for $t > T$ , see Eq. (20).}
\end{figure}

A vortex with a winding number 4 is thus created without
losing any atoms from the trap. It should be noted, however,
that it is technically difficult, albeit not impossible
\cite{mit}, to
introduce an optical plug with a few microns of radius along
the centre of the BEC whose radial dimension without optical plug
is of the order of a few microns. 

\section{Conclusions and Discussions}

The formation of a vortex in a BEC with $F=2$ in a Ioffe-Pritchard trap
has been considered by fully utilising the spinor degrees of
freedom. It was shown that a vortex with winding number 4
is created continuously from a condensate without a vortex,
by simply reversing the axial magnetic field $B_z(t)$.
This scenario has been studied with and without
an optical plug at the centre of the vortex. 
Some amount of the BEC is lost from the trap in the absence of
an optical plug while no atoms are lost in the presence of it.
Our numerical analysis shows that there remains a considerable
fraction of BEC even without the optical plug. The introduction
of an optical plug in a trapped BEC is difficult, albeit not
impossible. 

Our vortex has a large winding number 4 and is expected to be
unstable against decay into four singly quantised vortices
in the absence of an optical plug.
Whether a vortex with such a large winding number may be observable
depends on how large the lifetime of the metastable state is
compared to the trapping time of the BEC. Our prelimiary analysis
of the Bogoliubov equation suggests that the lifetime is of
the order of 100ms and these highly-quantised vortices exist for
a considerable duration of time. 

We would like to thank Kazushige Machida and Takeshi Mizushima
for discussions.
One of the authors (MN) 
thanks Takuya Hirano, Ed Hinds and Malcolm Boshier
for discussions. He also thanks Martti M. Salomaa for support
and warm hospitality in the Materials Physics Laboratory at
Helsinki University of Technology, Finland. MN's work
is partially supported by Grand-in-Aid from Ministry of Education,
Culture, Sports, Science and Technology, Japan (Project Nos.
11640361 and 13135215).
\vspace{1cm}\\
{\it Note Added} --- After we submitted our manuscript,
the MIT group reported the formation of vortices
according to the present scenario \cite{mit2}. They employed
hyperfine spin states $F=1$ and $F=2$ of $^{85}$Rb and
found that the vortex thus created had the winding number two in the
former case while four in the latter case, in consistent
with our prediction. The vortex state has considerably long lifetime,
at least 30ms after its formation, in spite of higher winding number,
which suggests that these vortices are rather stable. 
The stability analysis
of highly-quantised vortices is outside the scope of the present work and
will be published elsewhere.

We are grateful to Aaron Leanhardt for informing us of their result.

\end{document}